\documentstyle[emulateapj]{article}
\def\gsim{\;\rlap{\lower 2.5pt
 \hbox{$\sim$}}\raise 1.5pt\hbox{$>$}\;}
\def\lsim{\;\rlap{\lower 2.5pt
   \hbox{$\sim$}}\raise 1.5pt\hbox{$<$}\;}

\def\aal{{|\alpha|}}

\begin{document}
\title{On the Superradiance of Spin-1 Waves in an Equatorial Wedge
 around a Kerr Hole}
\author{Anthony N. Aguirre}
\affil{Department of Astronomy, Harvard University\\
MS 10, 60 Garden Street, Cambridge, MA 02138, USA\\
email: aaguirre@cfa.harvard.edu}
\submitted{Submitted to The Astrophysical Journal Letters}

\begin{abstract}
  Recently Van Putten has suggested that superradiance of magnetosonic
  waves in a toroidal magnetosphere around a Kerr black hole may play
  a role in the central engine of $\gamma$-ray bursts.  In this
  context, he computed (in the WKB approximation) the superradiant
  amplification of scalar waves confined to a thin equatorial wedge
  around a Kerr hole and found that the superradiance is higher than
  for radiation incident over all angles.  This paper presents
  calculations of both spin-0 (scalar) superradiance (integrating the
  radial equation rather than using the WKB method) and and spin-1
  (electromagnetic/magnetosonic) superradiance, in Van Putten's wedge
  geometry.  In contrast to the scalar case, spin-1 superradiance
  decreases in the wedge geometry, decreasing the likelihood of its
  astrophysical importance.

\noindent {\em Subject headings:} black hole physics -- gamma rays: bursts

\end{abstract}

\section{Introduction}
Van Putten (1999) has proposed that superradiant scattering of
magnetosonic waves by a Kerr black hole plays an important role in the
central engine of $\gamma$-ray bursts.  In his model, the tidal
breakup of a magnetized neutron star as it spirals into a Kerr hole
creates a massive torus and a toroidal magnetosphere.  Inside, there
is a current-free cavity which acts as a waveguide for fast
magnetosonic waves bouncing between the horizon and the torus.  As
shown by Uchida (1997), in the geometrical-optics limit these waves
(though not the Alfven waves) obey the same equations as the vacuum
electromagnetic waves (with `spin-weight' $s=1$) and hence are
amplified by `superradiant' scattering.  As the waves are assumed to
be perfectly confined to the cavity, this leads to an instability in
all superradiant modes.  To calculate the degree of superradiance, Van
Putten approximates the geometry as the interior of a very thin
equatorial wedge extending from the horizon to infinity (where the
torus lies).  This leads to a simple prescription for the angular
eigenvalues in the separated wave equation, and these can then be used
in the integration of the radial equation to obtain reflection and
transmission coefficients.  With this method Van Putten finds that
scalar waves ($s=0$) are reflected with a superradiance about ten
times that calculated in the full spheroidal geometry (Van Putten
1999; Press \& Teukolsky 1972).  This is notable because (in the full
geometry) superradiance increases with spin-weight, so this suggested
that magnetosonic ($s=1$) superradiance in the wedge may be more
efficient than either scalar superradiance in the wedge or $s=1$
superradiance over the full angular scale.  In this letter I calculate
the $s=1$ superradiance using a method analogous to Van Putten's and
find that the superradiance decreases, rather than increases, in the
`wedge' geometry.

\section{Method of Solution}

In the Newman-Penrose (1962) formalism, all field quantities are
represented by potentials obtained by projecting the fields onto a complex
tetrad of null-vectors $(l^\nu,n^\nu,m^\nu,m^{*\nu})$ which satisfy
$l_\nu n^\nu = 1 = -m_\nu m^{*\nu}$ and $m_\nu l^\nu = 0 = n_\nu
l^\nu.$ The electromagnetic field is then represented by the three
complex scalar potentials
$$\phi_0 \equiv F_{\mu\nu}l^\mu m^\nu,\ \ \phi_2 \equiv
F_{\mu\nu}m^{*\mu} n^\nu,$$
\begin{equation}
\phi_1 \equiv {1\over 2} F_{\mu\nu}(l^\mu n^\nu + m^{*\mu}m^\nu),
\end{equation}
 where $F_{\mu\nu}$ is the
EM field-strength tensor.  In a similar manner the Weyl and Ricci
tensors are expressed in terms of complex scalars, and the
Einstein-Maxwell equations are expressed in terms of these scalar
potentials.  See Chandrasekhar (1979) for a self-contained
treatment.

Teukolsky (1973) showed that, for a fixed Kerr geometry in
Boyer-Lindquist coordinates $(t,r,\theta,\phi)$, $\phi_0$ and $\phi_2$
(either of which contains the complete solution to the vacuum Maxwell
equations) yield separable solutions
\begin{equation}\phi_0 = e^{i\omega t}e^{im\phi}{_{\scriptscriptstyle 1}S_{m\omega}}(\theta){_{\scriptscriptstyle 1}R_{m\omega}}(r)\end{equation}
\begin{equation}{\bar \rho}^{2}\phi_2 = e^{i\omega t}e^{im\phi}{_{\scriptscriptstyle -1}S_{m\omega}}(\theta){_{\scriptscriptstyle -1}R_{m\omega}}(r)\end{equation}
where $\bar\rho = r-ia\cos\theta$ and $\omega$ is positive. The
equations for $R$ and $S$ are
\begin{eqnarray}
\label{eq-teukang}
\bigg[{1\over \sin\theta}{d\over d\theta}\sin\theta {d\over d\theta} &+& a^2\omega^2\cos^2\theta + 2 a \omega s \cos\theta \\ \nonumber
-{(m+s\cos\theta)^2\over\sin^2\theta}&+&E-s^2\bigg]{_{s}S_{m\omega}}(\theta; E)=0
\end{eqnarray}
and
\begin{eqnarray}
\label{eq-teukrad}
\bigg[\Delta^{-s}{d\over dr}\Delta^{s+1}{d\over dr} &+&
  {K^2+2is(r-M)K\over \Delta} -4ir\omega s \\ \nonumber
-E-2am\omega&-&a^2\omega^2+ s(s+1)\bigg]{_sR_{m\omega}(r;E)}. \nonumber
\end{eqnarray}
Here, $K\equiv (r^2+a^2)\omega+am$, $\Delta \equiv r^2+a^2-2Mr,$ $M$
and $a$ are the hole's mass and specific angular momentum, and
$E(m,\omega)$ is an angular eigenvalue.  The `spin-weight' $s$ takes
the value $s=+1$ if the equations are used for $\phi_0$, and $s=-1$
for $\phi_2$.  Scalar waves have $s=0$ and gravitational waves have
$s=\pm2.$

Chandrasekhar (1979) has shown how to cast the radial equation 
into a one-dimensional wave equation with potential barrier of the form
\begin{equation}
\left({d^2\over dr_*^2}+\omega^2\right)Z=VZ,
\ \ {d\over dr_*}={\Delta\over \rho^2}{d\over dr},
\label{eq-bareq}
\end{equation}
where $Z$ is some combination $Z = AR + B{d\over dr_*}R$ ($A$ and $B$
are functions of $r$ and $\theta$ given in C79; $B$=0 for $s=0$), and
where 
\begin{equation}\rho^2 = r^2+\alpha^2\ \ \ \ {\rm and}\ \ \ \ 
  \alpha^2=a^2+(am/\omega).
\end{equation} 
The new $r_*$ variable tends
to $+\infty$ as $r\rightarrow + \infty$, and tends to $\pm\infty$ as
$r\rightarrow r_+$ (the outer horizon).  The two signs are due to the
possible double-valuedness of the $r(r_*)$ relation.  For $0 < \omega
< -am/2Mr_+$ -- the `superradiant
interval' (Chandrasekhar 1979), the upper sign applies, and the potential can be
written
\begin{equation}
V={\Delta\over\rho^4}\left[\lambda + {\Delta\over\rho^4}|\alpha|(|\alpha|-4r)+2|\alpha|{r-M\over \rho^2}\right],
\end{equation}
where $\lambda = E+a^2\omega^2+2am\omega$ is a version of the angular
eigenvalue.  As $r\rightarrow r_{+}$, $r_* \rightarrow +\infty$, $V
\rightarrow 0$, and the solution tends to $Z\rightarrow e^{i\omega
  r_*}$ if one imposes the correct boundary condition at the horizon
(finite-amplitude ingoing waves as observed by infalling
observers; see Teukolsky 1973) and gives the wave unit amplitude.  The solution
may be integrated until it nears the singularity at $r=|\alpha|$,
where the differential equation in terms of $x\equiv
|\alpha|-r$ approaches
\begin{equation}
x^2{d^2Z\over dx^2}-x{dZ\over dx}=-{3\over 4}Z,
\label{eq-limde}
\end{equation}
the solution of which is\footnote{Not noted in Chandrasekhar 1979,
  this dangerously discards terms of $O(xZ)$, the order of the
  $x^{3/2}$ solution term.  But in the next-order expansion of
  eq.~\ref{eq-bareq}, a surprising cancellation justifies the
  procedure.}  $Z\rightarrow C_1 x^{3/2} + C_2 x^{1/2}$ Once $C_1$ and
$C_2$ are determined, the integration may be restarted with $r$
slightly greater than $|\alpha|$, with the solution $Z=iC_1 |x|^{3/2}
-iC_2 |x|^{1/2}.$ Note that the Wronksian $W_{r_*}[Z,Z^*]$ changes
sign at $r=\aal$.

As $r\rightarrow \infty$, $r_* \rightarrow \infty$ and $Z\rightarrow
C_{\rm inc}e^{i\omega r_*}+C_{\rm ref}e^{-i\omega r_*}$.  This allows
the definition of reflection and transmission coefficients
\begin{equation}
R = |C_{\rm ref}|^2|C_{\rm inc}|^{-2},\ \ \ T= |C_{\rm inc}|^{-2}
\end{equation}
which, due to the change in sign of the Wronskian, obey $R-T=1$
when $\omega$ is in the superradiant interval.  The numerical method
then entails integrating $Z$ from the boundary condition at $r
\rightarrow r_+$ out to $r\rightarrow \infty$ and finding $C_{\rm inc}$
and $C_{\rm ref}.$

The scalar case has a very similar potential, given by\footnote{This
  corrects the equation given in Chandrasekhar 1976.}
\begin{equation}
V={\Delta \over \rho^4}\left[\lambda + {1\over \rho^2}[\Delta + 2r(r-M)]
  -3{r^2\Delta \over \rho^4}\right].
\end{equation}
The scalar case can be integrated through the singularity using the
same method as employed for $s=1$, or by switching back to the usual
function $R(r)=Z(r_*)/\sqrt{|\rho^2|}$ and using
eq.~\ref{eq-teukrad} to integrate past the singularity.

These equations provide a prescription for computing the degree of
superradiant reflection given only the angular eigenvalue $E$
appropriate to the angular geometry and the $m$ and $\omega$ values.

\label{sec-nummeth}

\section{The Wedge Geometry}

Van Putten considers the simplified problem of a very narrow
equatorial wedge.  It is then assumed that the angular function is
constant across this wedge: $dS/d\theta=0$ in
equation~\ref{eq-teukang}.  Approximating then $\cos\theta \rightarrow
0$ and $\sin\theta \rightarrow 1$, eq.~\ref{eq-teukang} trivializes to
\begin{equation}
E \simeq m^2 + s^2.
\label{eq-trivang}
\end{equation}
This process is somewhat like creating an $l=0, m>0$ mode.  These
eigenvalues contrast with the usual eigenvalues of the `spin-weighted
spheroidal harmonics' (Teukolsky 1973) which follow from boundary
conditions of regularity at $\theta=0$ and $\theta=\pi$ and are given
(for small $a\omega$) by (Fackerell \& Crossman 1977):
\begin{equation}
E =l(l+1)+2a\omega{s^2m\over l(l+1)}+O[(a\omega)^2]
\label{eq-eigenapp1}
\end{equation}
for $s=1$, and for $s=0$ by
$$
E = l(l+1) + 2a^2\omega^2\left[{m^2-l(l+1)+{1\over2}\over(2l-1)(2l+3)}\right]+O[(a\omega)^4].
$$
For $s=0$, the wedge geometry changes the angular eigenvalue for $m=1$
from $E = 2+O[(a\omega)^2]$ to $E=1$.  The eigenvalue adds to the
height of the potential barrier, so the lower eigenvalue in the wedge
geometry leads to much {\em higher} superradiance in the scalar case
(see \S\ref{sec-results} below for numerical results).

An immediate worry arises in the $s=1$ case: the wedge geometry gives
$E=1+m^2$, which is {\em higher} than the minimal $l=|m|=1$ eigenvalues,
which are $<2$ (see eq.~\ref{eq-eigenapp1}; $m \omega < 0$ is required
for superradiance.)  This suggests that the result obtained for scalar
waves will not generalize to higher spins.  But before drawing this
conclusion firmly we must pose the problem as clearly as possible for
the magnetosonic waves.
\label{sec-eigenpot}

Van Putten's model postulates a force-free magnetosphere with a
current-carrying torus surrounding a current-free toroidal cavity
(see Van Putten 1999, fig. 2).  Since the boundaries of the region are
defined by magnetic field lines, component of $\vec B$ perpendicular
to the boundary must vanish there. The force-free condition implies
that $\vec E \perp \vec B$. This still leaves a choice of direction in
$\vec E$.  I shall choose {\em one polarization state}, in which the
components of $\vec E$ parallel to the boundary must vanish.  The
boundary conditions are, then, just like those at the boundary of a
perfect conductor.  In the wedge geometry, these are that $E_r =
E_\phi = B_\theta = 0$ near $\theta=\pi/2$, and hold in the rest-frame
of the matter in which the $\vec B-$field is anchored.  They also, it
turns out, hold in any frame connected to this frame by a boost in the
$\hat\phi$ direction, and therefore the conditions can be specified in
the `locally non-rotating frame' (LNRF; Bardeen 1972) as long as
there are predominantly $\phi$-direction bulk motions in the matter.

King (1977) gives the relevant LNRF field components explicitly in
terms of the Newman-Penrose potentials $\phi_i$, in the Boyer-Lindquist
coordinates. Evaluated at $\theta\rightarrow\pi/2$, the equations $E_r =
E_\phi = B_\theta = 0$ give
\begin{eqnarray}
\Im\left[\phi_2+{\Delta\over 2r^2}
\phi_0\right]=0 \\
\Im\left[\phi_2-{\Delta\over 2r^2}\phi_0\right]
-{2^{1/2}(r^2+a^2)\over ar}\Re[\phi_1]=0 \\
\Im\left[\phi_2-{\Delta\over 2r^2}\phi_0\right]
-{2^{1/2}a \Delta\over r(r^2+a^2)}\Re[\phi_1]=0.
\end{eqnarray}
Subtracting the last two implies that
\begin{equation}
\left({r^2+a^2\over a^2}-{r^2+a^2-2Mr\over r^2+a^2}\right)\Re[\phi_1]=0
\end{equation} 
Since its coefficient is always nonzero, we must have
$\Re[\phi_1]=0$, which then (using the first equation) implies that
the imaginary parts of $\phi_0$ and $\phi_2$ vanish.  So the necessary
and sufficient conditions for the proper field quantities to vanish at
$\theta=\pi/2$ are:
\begin{equation}\Im[\phi_0]=\Im[\phi_2]=\Re[\phi_1]= 0.\end{equation}

Because we have assumed a time dependence $\propto e^{i\omega t}$, in
specifying the value of the real or imaginary part of the complex
scalars, we must consider modes with frequencies $\pm\omega$;
likewise, with azimuthal mode numbers $\pm m.$ The radial
eigenfunctions (which, due to separation, do not change when adopting
the wedge-geometry) obey
\begin{equation}
{_sR_{m\omega}}(r;E)^* = {_sR_{-m-\omega^*}}(r;E^*),
\label{eq-rid3}
\end{equation}
therefore the solution, composed of two modes, of
\begin{eqnarray}
\label{eq-wedgesol}
\phi_0 &=& S(\theta){_{\scriptscriptstyle 1}R_{m\omega}}(r;E)e^{im\phi}e^{i\omega t} \\ \nonumber
&+&S(\theta){_{\scriptscriptstyle 1}R_{-m-\omega}}(r;E)e^{-im\phi}e^{-i\omega t}
\end{eqnarray}
has vanishing imaginary part
if $S(\theta)$, $\omega$ and $E$ are real.  Adopting then $S(\theta)=1$, the
`trivialized' angular equation will be satisfied, for $E=1+m^2.$
The angular solutions of $\phi_2$ are related to those of $\phi_0$ by~\cite{pt74}
\begin{eqnarray}
{_{\scriptscriptstyle -1}S_{m\omega}}&\propto&(\partial_\theta+m\csc\theta+a\omega\sin\theta) \\ \nonumber
&\times&(\partial_\theta+m\csc\theta+a\omega\sin\theta+\cot\theta){_{\scriptscriptstyle 1}S_{m\omega}} \nonumber
\end{eqnarray}
Multiplying this out and evaluating for $\theta\rightarrow\pi/2$ yields
\begin{equation}{_{\scriptscriptstyle -1}S_{m\omega}} \rightarrow ({\rm const.}) \times
[(m+a\omega)^2-1]{_{\scriptscriptstyle 1}S_{m\omega}},\end{equation}
so choosing both angular functions to be constant is consistent.
Since the radial solutions to for
$\phi_2$ also satisfy condition~\ref{eq-rid3}, this will lead to a
vanishing imaginary part of $\phi_2$ as well.  It can also be shown
(using Chandrasekhar 1979, eq. 7.186 for each mode) that $\Re[\phi_1]$ vanishes, so the full
boundary conditions are satisfied by eq.~\ref{eq-wedgesol}

Since the Maxwell equations are linear, we can evolve the two radial
solutions $R_{lm\omega}$ and $R_{l-m-\omega}$ independently from
$r=r_+$ to $r\rightarrow\infty$ to obtain the incoming and outgoing
wave amplitudes.  The amplitudes so obtained are invariant under
$m\rightarrow -m, \omega\rightarrow -\omega$, therefore the reflection
and transmission coefficients so obtained will be just those obtained
by considering either mode. This shows that the superradiance of
linearly polarized electromagnetic waves in a thin wedge between
perfect conductors can be calculated using of the equations outlined
in \S\ref{sec-nummeth}.  Note that a particular polarization has been
chosen (the other polarization state would have boundary conditions
which depend on the background magnetic field configuration), and
magnetosonic wave have been shown to coincide with linearly polarized
EM waves only in the geometrical-optics limits.  It is therefore
possible that magnetosonic superradiance will be different, but this
difference is unlikely to be large.


\section{Results and Discussion}

\begin{figure}
\plotone{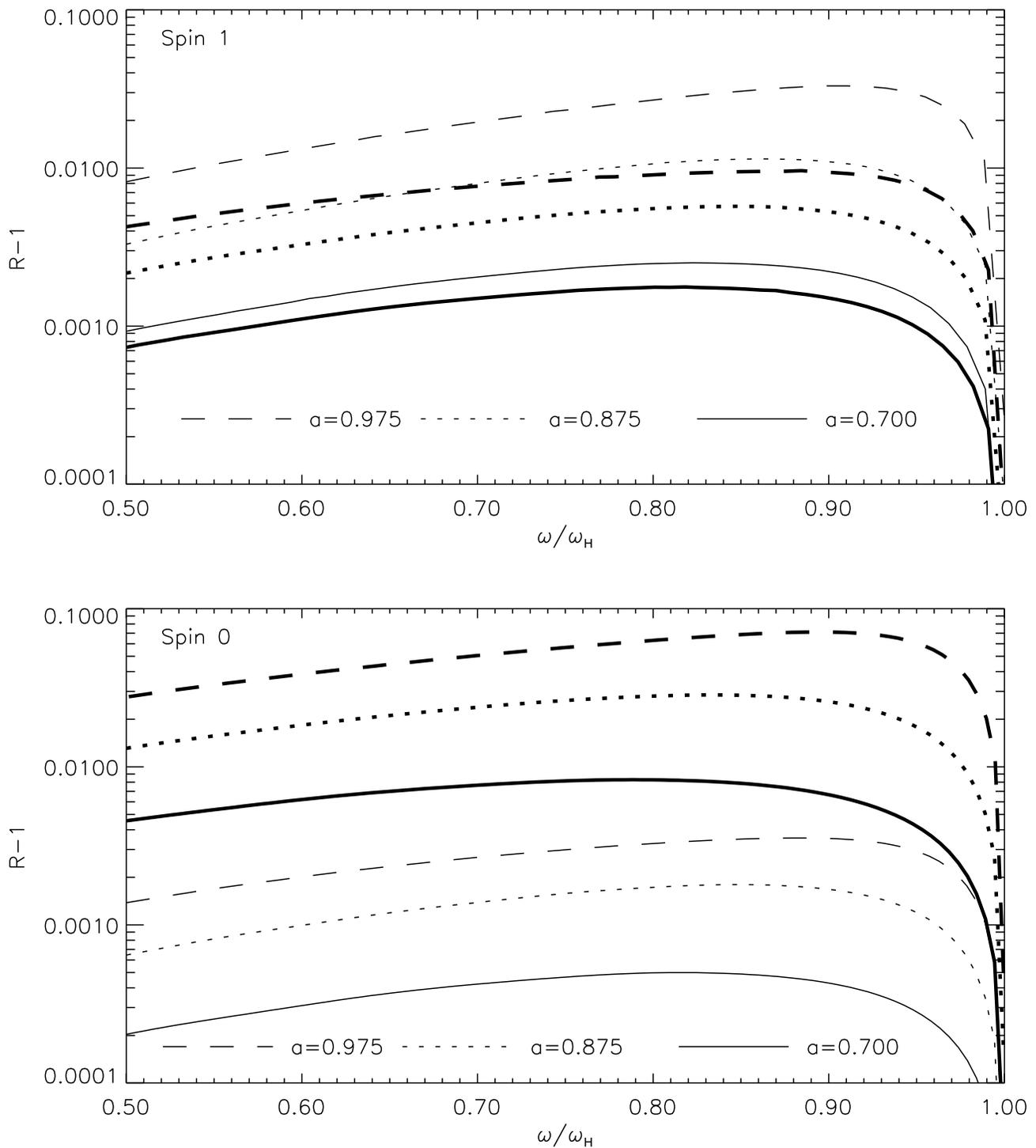}
\caption{{\bf Top:} Amplification $R-1$ for spin-1 
  (electromagnetic/magnetosonic) waves for various $a$, at frequencies
  normalized to $\omega_H \equiv -am/2Mr_+$.  Thin lines are with
  `usual' eigenvalues; thick lines use the `wedge approximation'
  $E=m^2+s^2$.  {\bf Bottom:} Same, for spin-0 (scalar) waves.}
\label{fig-spins}
\end{figure}

I have calculated the degree of superradiant reflection for
electromagnetic waves using both the usual eigenvalues as tabulated by
Press \& Teukolsky (1974), and using the eigenvalues for the
`wedge approximation' of $E=m^2+s^2$.  These are shown, in
fig.~\ref{fig-spins}, with $l=m=1$ and for various values of $a$, as
functions of $\omega.$ The maximum superradiance in the `usual'
electromagnetic case is $\approx 4.4\%$, and this falls to $\sim 1\%$
in the wedge approximation.  I have also computed the scalar wave
superradiance, also shown in fig.~\ref{fig-spins}.  Van Putten
employed the WKB approximation to estimate scalar superradiance, but
the potential varies over a scale comparable to the mode wavelength.
The $s=0$ results show that Van Putten's use of the WKB approximation
is not very accurate, and that in fact the scalar-wave superradiance
increases in the wedge geometry even more than he predicts, rising to
a maximum value of $\sim 7\%$, from a maximum of $\sim 0.3\%$ in the
full geometry.

The somewhat counter-intuitive result that superradiance decreases in
the wedge geometry for $s=1$ while increasing for $s=0$ can be
understood using the following heuristic argument.  The angular
eigenvalue $E$ links the angular and radial parts of the wave
equation, effectively adding a term to the potential representing the
angular momentum barrier; this situation is familiar from
quantum mechanics, where (as in the $\omega=0$ case here), $l(l+1)$ gives
the total angular momentum ${\cal L}^2$ associated with the
eigenfunction labeled by $(l,m)$. Also familiar from quantum
mechanics,
$${\cal L}^2 = \langle \vec L^2\rangle = \langle L_x^2 \rangle +
\langle L_y^2 \rangle + \langle L_z^2 \rangle \ge \langle L_z^2
\rangle = m^2,$$
i.e. $m^2$ is a lower limit to the total angular
momentum (resulting from the azimuthal variation), regardless of the
value of $l$ (which is, of course, always $\ge m$ in the full-sphere
case).  Generalizing this to $s \ge 0$, it is possible to construct an
operator $K_r$, analogous to $L_z$ but with eigenvalue $s$,
representing the angular momentum (helicity) about the radial
direction, rather than about the $\hat z$
direction (Goldberg et al. 1967; Campbell 1972). For $\theta=\pi/2$, the radial
direction lies in the $\hat x-\hat y$ plane, and we can write
\begin{equation}
{\cal L}^2=\langle L_x^2 \rangle +
\langle L_y^2 \rangle + \langle L_z^2 \rangle = L_z^2 + K_r^2 = m^2+s^2
\end{equation}
These equations imply that the total angular momentum has a value
$m^2+s^2$ when the wave is confined to $\theta=\pi/2$, in agreement
with the `trivialized' angular equation~\ref{eq-trivang}.
That is, the helicity of the $s>0$ wave provides an extra component of
the angular momentum barrier in the wedge which is not present in the
scalar case.

The phenomenon of superradiant scattering from a Kerr hole has been
understood for thirty years, but has yet to find astrophysical
applications because the degree of amplification for electromagnetic
waves tends to be small; creation of an instability requires a very
efficient `mirror' with reflectivity of $\gsim 95\%$.  Van Putten only
assumes $0.5-5\%$ superradiance in calculating timescales in his
model, but his interesting analysis of the thin equatorial wedge
suggested that the wedge geometry might greatly enhance superradiance,
making its astrophysical importance very plausible.  Unfortunately the
present, more detailed, calculations do not bear out this idea.  The
cavity in which the magnetosonic waves are confined must be $\sim
99\%$ dissipation-free to create an instability; the assumption that
such cavities can form in a natural setting requires justification.
Moreover, energy leaking to higher-$m$ or higher-$\omega$ modes will
be even more weakly amplified.
A remaining possibility for the importance of superradiance in a less
idealized setting remains, however.  If the wave is reflected (or
perhaps trapped) close to the resonance radius $r=\aal$ (rather than
at infinity), the outer part of the potential barrier could be avoided
and superradiance increased.  Whether this increase might be
sufficient to be astrophysically relevant requires further analysis.
\label{sec-results}

\acknowledgements

I thank Ramesh Narayan, Lars Hernquist, George Rybicki, Bill Press and
Maurice Van Putten for useful discussions.  This work was supported in
part by the National Science Foundation grant no. PHY-9507695.

\end{document}